\documentclass[aps,pre,twocolumn,floatfix,nofootinbib,amsmath,amssymb]{revtex4}

\usepackage[english]{babel}
\usepackage{graphicx}
\usepackage{textcomp}
\usepackage{amsmath}
\usepackage[usenames,dvipsnames]{xcolor}

\begin{document}
	
	\title{A simple statistical physics model for the epidemic with incubation period  }
	\author{David B. Saakian$^{1}$}
	

	\affiliation{$^1$
		A.I. Alikhanyan National Science Laboratory
		(Yerevan Physics
		Institute) Foundation,\\
		2 Alikhanian Brothers St., Yerevan 375036, Armenia}

	\date{\today}

	\begin{abstract}
{
Based on the classical SIR model, we derive a simple modification for the dynamics of epidemics with a known incubation period of infection. The model is described by a system of integro-differential equations. Parameters of our model directly related to epidemiological data. We derive some analytical results, as well as perform numerical simulations. We use the proposed model to analyze COVID-19 epidemic data in Armenia. We propose a strategy: organize a quarantine, and then conduct extensive testing of risk groups during the quarantine, evaluating the percentage of the population among risk groups and people with symptoms.
}		
\end{abstract}

	\maketitle
\section{Introduction}

Mathematical modeling for epidemiology has a rather long history, 
dating back to the studies by D. Bernoulli [1]. Later,  Kermack and McKendrick [2] proposed their prominent theory for infectious disease dynamics, which influenced the following SIR and related models. By the end of the last century, significant progress in the field was made (a systematic literature review for this period is presented in Anderson and May's book [3]). 
The COVID-19 pandemic has drawn the attention of researchers from all over the world and different areas to epidemic modeling. One of the simplest SIR models for the virus spread in Northern Italy was introduced in \cite{ga}. Another research group used the logistic equation to analyze empirical data on the epidemic in different states \cite{did}.

Here, we mainly focus on mean-field models that discard the spatial dependence of the epidemic process. Therefore, we avoid network models of epidemics.  Moreover, it is crucial to consider the final incubation period of the disease to construct a correct model for the COVID-19 case. Taking into account this distinctive feature, we consider the dynamics of the aged-structured population, which a well-known problem in evolutionary research \cite{ha} - \cite{sa10}. Generally, epidemic models have a higher order of non-linearity than evolutionary models, although there are some similarities between these two classes.

In this study, we derive a system of integro-differential equations based on the rigorous master equation that adequately describes infection dynamics with an incubation period, e.g., COVID-19.  First, we discuss the SIR model. Then, we move on to its modification and apply it to the data on the COVID-19 epidemic in Armenia. 

Consider the SIR model where the parameter $S$ stands for the number of susceptible people, $I$ for the number of infected people, and $R$ for the number of people  who have recovered and developed immunity to the infection. We assume that  $S, I, R$ satisfy the constraint $S+I+R=N$.
\begin{eqnarray}
\label{e1}
&&\frac{dS}{dt}=-a SI,\nonumber\\
&&\frac{dI}{dt}=aSI-b I,\\
&&\frac{dR}{dt}=bI,\nonumber
\end{eqnarray}
where $1/b$ is the period when the infected people are contagious.

The parameter $a$ can be obtained from the empirical data on infection rate:
\begin{eqnarray}
\label{e2}
&&a=\frac{\alpha}{N},\nonumber\\
&&N=S+I+R.
\end{eqnarray}
Thus, we assume that a healthy person is infected with a probability
 proportional to the fraction of infection in the population.
Probability is also proportional to the population density. 

One of the most widely discussed and crucial parameters in epidemiological data is the basic reproduction number of the infection:
\begin{eqnarray}
\label{e3}
R_0=\frac{\alpha}{b}.
\end{eqnarray}
For the COVID-19, it has been estimated as \cite{ga}:
\begin{eqnarray}
\label{e3a}
2<R_0<4\nonumber
\end{eqnarray}

In fact, the real data allows us to measure three main parameters: the exponential growth coefficient at the beginning of the epidemic; the minimum period of time, in which
an infected person can transmit the infection; and the maximum period, when an infected person ceases to transmit the infection.

The most important objectives of the investigation are the maximal possible proportion of the infected population, and then the period before the peak of the epidemic.

\section{SIR model with incubation period}

Consider the spread of infectious diseases with a recovery period up to $T$ days. At the $t$-th moment of time, we have $ S (t) $ for the size of the susceptible population, $R (t)$ for the recovered population. We divide the infected population according to the age of infection, looking at time intervals $\delta$ and defining $ I_l (t) $ as the number of infected people
with the age of infection in $ (l \delta, (l + 1) \delta) $.
We assume that the incubation period for a random infected person is $L$ and the infection spreads from $L$ to $T$ days.
Below, we take $ \delta \to 0 $  for the continuous mode of time limit.

Assuming that the spread of infection has a rate $a$, we obtain the following system of equations:
\begin{eqnarray}
\label{e4}
&&\left(S(t+\delta)-S(t)\right)/\delta =-a S(t)\sum_L^TI_l(t)\nonumber\\
&&I_l(t+\delta)=I_{l-1}(t),\nonumber\\
&&R(t+\delta)=R(t)+\delta I_{(T/\delta)}(t),\nonumber\\
&&I_0(t+\delta)=a\delta  S(t)\sum_L^T I_{l}(t),
\end{eqnarray}
where the coefficient $a$ is expressed via the infection rate coefficient $A$,
\begin{eqnarray}
\label{e5}
a=\frac{A}{N}.
\end{eqnarray}

We suggested that after $ T $ days a person recovers and  the patients are not isolated from the rest of the population between days $ L $ and $ T $.
Eq. (\ref{e4}) describes the dynamics of the population over discrete time, which is the right choice for numerical simulation.

Consider now the continuous-time version of the model. In the limit of small $ \delta $, we introduce the continuous function $ I_l (t) = I (x, t) $, where 
$ \delta I (x, t) $ is the size of the infected population with age $ x, x + \delta $.
The continuous time versions for the first three equations are:
\begin{eqnarray}
\label{e6}
&&dS/dt=-a S(t)\int_L^TdxI(x,t)\nonumber\\
&&dI/dt=-dI/dx,\nonumber\\
&&dR/dt=I(T,t).
\end{eqnarray}
The solution of the second equation in Eq.(1) is
\begin{eqnarray}
\label{e7}
I(x,t)=J(t-x),
\end{eqnarray}
where we denote $J(x)=I(0,t)$

Let us look at the difference
\begin{eqnarray}
\label{e8}
&&J(t+\delta)=a  S(t)\int_L^T dxJ(t-x),\nonumber\\
&& J(t)=a  S(t-\delta)\int_L^T dxI(t-x-\delta).
\end{eqnarray}
Using the latter expressions, we get the following full system of equations:
\begin{eqnarray}
\label{e9}
&&dS/dt=-a S(t)\int_L^TdxJ(t-x),\nonumber\\
&&\frac{dJ(t)}{dt}\nonumber\\
&&=a \frac{dS}{dt}\int_L^T dxJ(t-x)+aS(t)\left[J(t-L)-J(t-T)\right],\nonumber\\
&&dR/dt=J(t-T).
\end{eqnarray}
At the start, when $S'\ll S$
\begin{eqnarray}
\label{e9a}
\frac{dJ(t)}{dt}=aS(0)\left[J(t-L)-J(t-T)\right].\nonumber
\end{eqnarray}
Substituting an ansatz $J(t)=e^{k t}$, we get: 
\begin{eqnarray}
\label{e10}
1=a  S(0)\frac{e^{-k L}-e^{-k T}}{k}.
\end{eqnarray}
At $k\to 0$, we get:
\begin{eqnarray}
\label{e10a}
1=a  S(0) (T-L).\nonumber
\end{eqnarray}
For increasing $\alpha$, we get an increasing value of $k$ as well.
In the SIR model the epidemic threshold is at $R=1$ or $a=b$, so  our model is similar to SIR with $b=1/(T-L)$.
 
 In Fig 1., we analyze the epidemiological data for COVID-19 in Armenia using our model. We examine the dynamics of infected population in Armenia from March 25, when the quarantine in the country has been introduced by the government, until April 5.

 \begin{figure}[h!]
 	\center
 	\includegraphics[scale=0.55]{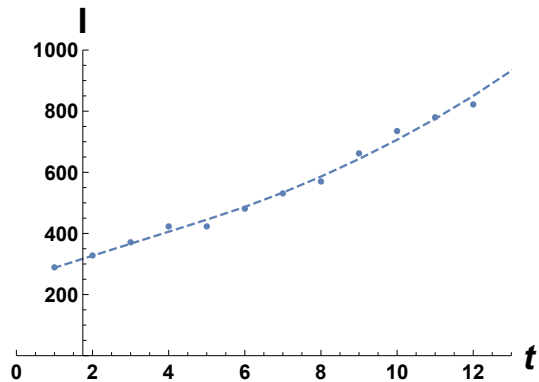}
 	\caption{ 
 		The infected population size $I(t)$ in Armenia with respect to time in days. The count of days starts at March 24.
 		The infected population fraction $p$ is shown versus $t$. The total population size is 3 million. Here, 
 		$k=0.0887,L=5,T=15,a=0.235$. At the start, we have 250 infected people.
 	}
 \end{figure}

\section{General case of our model}

Let us consider the case, when the infectivity (the ability to transfer the infection to  susceptible individuals) of infected individuals depends on the age of infection (via a kernel $f(x)$), also the population with the age $x$ is diluted with the rate $g(x)$. The latter seems to be a reasonable assumption, since an infected individual with a large age reveals some symptoms of infection, therefore, has chances to be isolated.  
Now Eq. (4) is modified:
\begin{eqnarray}
\label{e11}
&&(S(t+\delta)-S(t))/\delta =-a S(t)\sum_L^TI_l(t)f(l \delta),\nonumber\\
&&I_l(t+\delta)=I_{l-1}(t)-I_{l}(2)g(l\delta),\nonumber\\
&&R(t+\delta)=R(t)+\delta I_(T/\delta),(t)\nonumber\\
&&I_0(t+\delta)=a\delta  S(t)\sum_L^T I_{l}(t)f(l\delta). 
\end{eqnarray}
The continuous time limit gives the following system of equations:
\begin{eqnarray}
\label{e12}
&&dS/dt=-a S(t)\int_L^TdxI(x,t)f(x),\nonumber\\
&&\frac{dI(0,t)}{dt}\nonumber\\
&&=a \frac{dS}{dt}\int_L^T dxI(x,t)+aS(t)\int_L^TdxI'_t(x,t)f(x), \nonumber\\
&&dR/dt=I(T,t)+\int_L^TdxIt(x,t)g(x),\nonumber\\
&&\frac{dI(x,t)}{dt}=-\frac{dI(x,t)}{dx}-g(x)I(x,t).
\end{eqnarray}
Consider now the asymptotic solution:
\begin{eqnarray}
\label{e13}
I(x,t)=q(x)e^{kt}.
\end{eqnarray}
Then, we get the following equations:
\begin{eqnarray}
\label{e14}
q(x)=e^{-\int_0^x(k+g(y))dy},
\end{eqnarray}
and
\begin{eqnarray}
\label{e15}
1=a\int_L^Tdxf(x)e^{-\int_0^x(k+g(y))dy}.
\end{eqnarray}
Thus, we derive for the epidemics threshold:
\begin{eqnarray}
\label{e16}
1=a\int_L^Tdxf(x)e^{-\int_0^xg(y)dy}.
\end{eqnarray}

\subsection{The specific functions $g(x)$}
Let us analyze our Eq. (16). If we are trying to reduce the growth rate $k$, it can be done in two ways:
\begin{enumerate}
\item reducing the number of contacts, $A$,

\item increasing the $g(x)$ via containment activities.
\end{enumerate}

Let us introduce non-zero reduction, just after 7 days, $g(x)=g$. Hence, we get the following result:
\begin{eqnarray}
\label{e17}
1=a\left[\int_L^{T_1}dxe^{-kx}+\int_{T_1}^{T_2}dxe^{-(k+g)x}\right].
\end{eqnarray}
We should estimate the value of $g$ that stops the epidemics.
\section{Armenian case}

We now apply the generalized version of the model to the epidemics  in Armenia.
We look at two periods of epidemics: the first period from March 24 to April 5 and the later (second period), when quarantine starts to work efficiently.

\subsection{The choice g=0 in the model.}
Let us first take $g=0$ for the first period, see Table 1.
\begin{center}
	\begin{table}[tbhp]
		\caption{ The model from section III for different values of parameter $A$. We choose the parameter $A$ from the actual data for the exponential growth rate $k$. The third row defines the epidemic threshold }
		\begin{tabular}{l@{\hspace{4mm}}c@{\hspace{8mm}}c@{\hspace{6mm}}c}
			\hline\hline A  &  g &   k   &   \\
			[0.5ex] \hline
			0.235 & 0 & 0.088  &   \\
			0.135 & 0& 0.031& \\
			0.1 & 0& 0.0& \\
			\hline\hline
		\end{tabular}
	\end{table}
\end{center}

The parameter $A$ in our model is proportional to the number of human contacts during the day.
In the first time period, we have had $a=0.235, k=0.088$.
After the quarantine in Armenia, $k$ decreased from the value $0.088$ till the value $0.031$, with $a=0.135$. The critical value of $A$ to eliminate the epidemics is $a=0.1$.
We reduced $43 \% $ of human contacts. The  reducing further $26 \%$ of remaining contacts, we can eliminate the epidemics.

Let's evaluate what degree of $g$ we need to eliminate the epidemic at given values ​​of$ A$, if we attribute the current situation to $ g = 0 $. For the case of quarantine, we take the current value $ k = 0.031 $, then we see that $g = 0.035$ eliminates
epidemics.

The parameter $A$ in our model is proportional to the number of human contacts during the day.
In the first period of epidemics, we had $ a = 0.235, k =  0.088.$
After quarantine in Armenia, $k$ decreased from $0.088$ to $0.031,$ with $ a = 0.135 $. The critical value of $A$ for the elimination of epidemics is $ A = 0.1 $.
We have reduced $43\%$ 
of human contacts. A further reduction of $26\%$
 of the remaining contacts, we can eliminate the epidemic.

For the  case without quarantine,  $k=0.088$, we need $g=0.12$ to eliminate the epidemics, much more efforts compared to the previous considered case.

\section{ The choice g=0.05 at the first period of epidemics}.

\begin{center}
	\begin{table}[tbhp]
		\caption{ The model from section III for $g=0.05$ different values of parameters. From the actual data for the exponential growth rate $k$, we choose the parameters $A$. The third line defines the epidemic threshold }
		\begin{tabular}{l@{\hspace{4mm}}c@{\hspace{8mm}}c@{\hspace{6mm}}c}
			\hline\hline A  &  &   k   &   \\
			[0.5ex] \hline
			0.333 &  & 0.088  &   \\
			0.2   &  & 0.031& \\
			0.151 & & 0.0& \\
			\hline\hline
		\end{tabular}
	\end{table}
\end{center}

Let us take $g=0.05$ for the first period (we identify the $35 \%$ of infected individuals during a week), see Table II. Then,  we apply $A=0.333$  for the first period, $A=0.2$ during the second period and we need  $A=0.151$ to eliminate the epidemic.
Due to the quarantine, we reduced $40 \%$  of contacts, we now needs in reducing of  $25 \%$  of existing contacts.  Holding current values of contacts, we need rising the value of $g$ from $0.05$ to $0.09.$

 We verified that taking $g=0.15$ before the quarantine does not give adequate results. 
How can we increase $g$ in practice? Testing the $20 \%$  per week in  high risk groups of the population, we can eliminate the epidemics.

\section{Conclusion}

In this paper, we introduced a version of SIR model for infection spreading with known incubation period. This model was applied to analyze the COVID-19 epidemic data in Armenia. We constructed the simplest version of population dynamics of age-structured population. Close work has been done in \cite{kr}, which is related to SIR model. In \cite{kr}, a temporal kernel $F(t)$ has been introduced that modulates the infectivity of each infected individual.
Compared to such model, we introduced the distribution of infected population at given moment of time via an age of infection, instead of looking just long history of focus populations.
In other works related to the population dynamics of age-structured population,  the differential equations with time delay usually have been considered. Instead, we use integro-differential system of equations, which seems to be an adequate approach to the current situation with COVID-19 epidemic.

From our perspective, the proposed approach significantly changes the epidemiological picture (compare to classical SIR models),  since the virus is active for about two weeks.
Next, we introduced two functions: $f (x)$, which describes the distribution of infectivity by age, and $g (x)$, which describes the content measures.
In the normal SIR model, we have two parameters for the rate of infection and the removal of infections.
In our integro-differential model, mapping to elementary processes is straightforward: we just need the velocity parameter $a$ and two periods: the incubation period $L$ and recovery period $T$ with symptoms after the carrier is separated from society.
We derive an analytical result for exponential growth in the early stages of epidemics, as well as for the epidemic threshold. It will be very interesting to investigate the transitional situation near the threshold. We suggest simply making numbers and choosing a parameter value to match the correct exponential growth.

We applied our model to understand the situation with epidemics in Armenia.
What is  advantageous in our model, that we can clearly separate two aspects of the epidemic: contact strength (through coefficient $A$) and deterrence measures through parameter $g$.
We check that in fact we need to make minimal efforts to stop the epidemics, and testing is much cheaper during quarantine.
Currently, if we detect only 3.5\% of the infected population per day, strictly monitoring the symptoms, we can stop the epidemic.
This is much more complicated without quarantine.

I thank Armen Allahverdyan, Pavel Krapivsky, Ruben Poghosyan, Didier Sornette,  and Tatiana Yakushkina for useful discussions.
The work is supported by the Russian Science Foundation under grant 19-11-00008 from Russian University of Transport

\end{document}